\documentclass[nofootinbib,reprint,superscriptaddress]{revtex4-2}

\makeatletter 
    
\renewcommand\onecolumngrid{
\do@columngrid{one}{\@ne}%
\def\set@footnotewidth{\onecolumngrid}
\def\footnoterule{\kern-6pt\hrule width 1.5in\kern6pt}%
}

\renewcommand\twocolumngrid{
        \def\footnoterule{
        \dimen@\skip\footins\divide\dimen@\thr@@
        \kern-\dimen@\hrule width.5in\kern\dimen@}
        \do@columngrid{mlt}{\tw@}
}%

\makeatother    

\usepackage[bottom]{footmisc}
\usepackage{amsmath,amssymb,marginnote}
\usepackage{graphicx,subfigure}
\usepackage{dcolumn}
\usepackage{bm}
\usepackage[mathlines]{lineno}
\usepackage[colorlinks=true, allcolors=blue]{hyperref}
\usepackage[usenames]{color}

\newcommand\barparent[1]{\overset{%
    \scriptscriptstyle(-)}{#1}}
\begin{document}
\title{Muon-induced collisional flavor instability in core-collapse supernova}
\author{Jiabao Liu}
\affiliation{Department of Physics and Applied Physics, School of Advanced Science \& Engineering, Waseda University, Tokyo 169-8555, Japan}
\author{Hiroki Nagakura}
\affiliation{Division of Science, National Astronomical Observatory of Japan, 2-21-1 Osawa, Mitaka, Tokyo 181-8588, Japan}
\author{Ryuichiro Akaho}
\affiliation{Department of Physics and Applied Physics, School of Advanced Science \& Engineering, Waseda University, Tokyo 169-8555, Japan}
\author{Akira Ito}
\affiliation{Department of Physics and Applied Physics, School of Advanced Science \& Engineering, Waseda University, Tokyo 169-8555, Japan}
\author{Masamichi Zaizen}
\affiliation{Faculty of Science and Engineering, Waseda University, Tokyo 169-8555, Japan}
\author{Shun Furusawa}
\affiliation{College of Science and Engineering, Kanto Gakuin University, Kanagawa 236-8501, Japan}
\affiliation{Interdisciplinary Theoretical and Mathematical Sciences Program (iTHEMS), RIKEN, Wako, Saitama 351-0198, Japan}
\author{Shoichi Yamada}
\affiliation{Department of Physics, School of Advanced Science \& Engineering, Waseda University, Tokyo 169-8555, Japan}
\affiliation{Research Institute for Science and Engineering, Waseda University, Tokyo 169-8555, Japan}

\begin{abstract}
Neutrinos are known to undergo flavor conversion among their three flavors. In the theoretical modeling of core-collapse supernova (CCSN), there has been a great deal of attention to recent discoveries of a new type of neutrino flavor conversions, namely collisional flavor instability (CFI), in which the instability is induced by the flavor-dependent decoherence due to the disparity of neutrino-matter interactions among flavors. In this paper, we study how the appearance of on-shell muons and associated neutrino-matter interactions can impact CFIs based on linear stability analysis of flavor conversions. Some striking results emerge from the present study. First, we analytically show that breaking beta- and pair equilibrium is a necessary condition to trigger CFIs. This also indicates that CFIs with on-shell muons could appear in $e \tau$ and $\mu \tau$ neutrino mixing sectors in very high-density region ($\gtrsim 10^{13} {\rm g/cm^{3}}$), exhibiting a possibility of large impacts of CFIs on CCSN. Second, resonance-like CFIs, having a much higher growth rate than normal CFIs, can be triggered by muons. The resonance point of CFIs is different between $e \tau$ and $\mu \tau$ sectors; the former (latter) occurs at $\mu_{e (\mu)} = \mu_{n} - \mu_{p}$, where $\mu_{i}$ denotes the chemical potential of $i$ constitute ($n$ and $p$ represent neutrons and protons, respectively). Our result suggests that the non-linear evolution of CFI with on-shell muons would induce flavor conversions with the complex interplay among all three different neutrino-mixing sectors.

\end{abstract}
\maketitle

\section{Introduction}

The life of stars with mass $\gtrsim10\, M_\odot$ culminates in a dramatic explosion known as core-collapse supernovae (CCSNe). Such catastrophic astrophysical events begin when the iron core of the star meets critical conditions to trigger electron capture and/or photodissociation of heavy nuclei. These reactions reduce matter pressure, making the iron core more compact, and facilitating these reactions more vigorous, thus establishing a gravitational instability. During the gravitational collapse, electron-type neutrinos ($\nu_e$) are abundantly produced through charged-current interactions, which cools the core and accelerates the collapse further. When the baryon density in the inner core goes over that of nuclear saturation one, the nuclear force suddenly dominates matter pressure, which makes the core bounce back, creates a shock wave, and leaves behind a proto-neutron star (PNS) at the center. In the post-bounce phase, other types of neutrinos, including electron-type anti-neutrinos ($\bar{\nu}_e$), muon/tau neutrinos, and their anti-partners ($\barparent{\nu}_\mu/\barparent{\nu}_\tau$), can also be abundantly produced via multiple weak interaction processes. These neutrinos play crucial roles in PNS cooling and also transfer energy from the PNS core/envelope to the vicinity of the shock wave. Although most of these neutrinos freely escape from the post-shock region, some of $\barparent{\nu}_e$ are captured (or absorbed) by nucleons via charged-current processes. This process works to deposit the neutrino energy into the post-shock matter, which pushes the shock wave to the outer radii. According to recent detailed CCSN simulations, such neutrino-driven explosions aided by multi-dimensional fluid instabilities have been observed rather commonly for a wide range of progenitor mass; see, e.g., (\cite{Ott_2018,10.1093/mnras/stz3223,10.1093/mnras/staa1048,Harada_2020,Bollig_2021,10.1093/mnras/stac1586,PhysRevD.107.043008}).

Although there is emerging a consensus among different CCSN groups that the delayed neutrino heating mechanism potentially accounts for a majority of CCSN explosions, we should keep in mind that there are great uncertainties in these modeling. In dense neutrino environments such as CCSN cores, neutrino self-interactions can induce collective neutrino-flavor conversions. Due to the non-linear nature of self-interactions, the dynamics of flavor conversions are qualitatively different from those in vacuum and medium \cite{PANTALEONE1992128}. One of the most striking phenomena may be fast neutrino-flavor conversion (FFC), which is associated with fast-pairwise flavor instability. Flavor conversions could occur in timescales of picoseconds, which could drastically alter the neutrino flavor structures~\cite{PhysRevD.46.510, Sigl1993GeneralKD, doi:10.1146/annurev.nucl.012809.104524, CHAKRABORTY2016366, doi:10.1146/annurev-nucl-102920-050505, Richers2020, PhysRevD.107.043024, PhysRevD.107.123024}. FFC has been increasingly observed in multi-dimensional CCSN simulations \cite{PhysRevD.100.043004, Nagakura_2019, PhysRevD.101.023018, PhysRevD.101.043016, PhysRevD.101.063001, Abbar_2020, PhysRevD.103.063013, PhysRevD.104.083025, Harada_2022, Akaho_2023, PhysRevD.99.103011, PhysRevD.101.023018, PhysRevD.101.043016, PhysRevD.100.043004, PhysRevD.104.083025, Harada_2022, Akaho_2023}, motivating detailed studies into its non-linear evolution \cite{PhysRevD.101.043009, PhysRevD.102.103017, PhysRevD.104.103003, PhysRevLett.128.121102, PhysRevD.104.103023, PhysRevD.106.103039, PhysRevLett.129.261101, PhysRevD.107.103022, PhysRevD.107.123021, xiong2023evaluating}, and its interplay with neutrino-matter collisions \cite{PhysRevD.103.063001, PhysRevD.105.043005, PhysRevD.106.043031, PhysRevD.103.063002, Kato_2021, 10.1093/ptep/ptac082, PhysRevD.105.123003, Kato_2022, PhysRevD.106.103031, PhysRevD.103.063002, PhysRevLett.122.091101}. Recent studies show that the neutrino radiation fields affected by FFC in the CCSN core differ remarkably from those modeled by classical neutrino transport~\cite{PhysRevD.107.063025, PhysRevLett.130.211401, nagakura2023basic}. This suggests that fluid dynamics, nucleosynthesis, and observed neutrino signals could be significantly influenced by FFC.

The recently discovered~\cite{PhysRevLett.130.191001} and now well-established flavor instability
~\cite{PhysRevD.108.083002,liu1,PhysRevD.107.083034,PhysRevD.106.103029,PhysRevD.106.103031,PhysRevD.109.063021}, namely collisional flavor instability (CFI), is another noticeable channel to induce large flavor conversions in CCSN cores. It has also been found recently that there are resonance-like features in CFI, which only happens when electron-neutrino lepton number (ELN) nearly equals that for heavy-neutrinos (XLN). Although the spatial region of the resonance-like CFI would be very narrow~\cite{PhysRevD.108.083002,CFIakaho}, the growth rate of CFI can be enhanced substantially. On the other hand, there is another distinct feature of CFI from FFI. It can occur in regions where angular distributions of all species of neutrinos are isotropic. This suggests that CFI might occur in deeper regions of the CCSN core compared to FFI. We thus anticipate that CFI potentially has considerable effects on CCSN dynamics~\cite{liu2,CFIakaho,PhysRevD.107.083016,shalgar2023neutrinos} (and also on binary neutron star mergers~\cite{PhysRevD.109.043046}).

In this paper, we study a new aspect of CFI: the impacts of muonization. According to previous studies~\cite{PhysRevLett.119.242702,PhysRevD.102.123001}, on-shell muons may appear during the pre-explosion phase, which potentially has an influence on CCSN dynamics~\cite{PhysRevLett.119.242702,PhysRevD.102.123001}. 
A crucial point here is that the appearance of muons opens various weak interaction channels~\cite{PhysRevD.102.023037,10.1093/ptep/ptac118}, which have significant impacts on both neutrino radiation fields and fluid dynamics. This also suggests, but has not been even investigated so far, that CFI can be qualitatively different from those without muons. In fact, weak interaction processes involving on-shell muons have been neglected so far in previous CFI studies. It should also be mentioned that collision terms between $\nu_{\mu}$ and $\nu_{\tau}$ are significantly different from each other, indicating that heavy-leptonic neutrinos can no longer be treated collectively. This indicates that CFIs could occur in multiple neutrino-mixing sectors of $e \mu$, $e \tau$, and $\mu \tau$. These new features of CFI are worth investigating.

In the next section (Sec.~\ref{sec:PPM}), we summarize our method to study CFI with on-shell muons. After introducing essential information on linear stability analysis of CFI in Sec.~\ref{sec:CFI}, we present some novel features of CFI in Sec.~\ref{sec:results}. Finally, we conclude the present study with the summary of our findings in Sec.~\ref{sec:conclusion}.

\section{Method}
\label{sec:PPM}

The self-consistent way to analyze CFI in CCSN environments is to carry out stability analyses of CFI with respect to neutrino radiation fields obtained by realistic CCSN simulations, as done in our previous studies~\cite{liu2,CFIakaho}. However, we do not currently have CCSN models with muonization taken into account.

We also note that the appearance of muons depends on many factors~\cite{PhysRevLett.119.242702}. These complexities involved in CCSN simulations make it hard to inspect the intrinsic properties of CFI with muons. For this reason, we take another approach in this study. We employ a fluid profile from one of our CCSN models without muon and then carry out stability analyses by systematically changing muon fraction ($Y_{\mu}$) in a parametric manner (see Sec.~\ref{subsec:backmod}). Below we describe the procedure in detail.

\subsection{CCSN model}\label{subsec:CCSNmodel}

\begin{figure}[ht]
    \centering
    \includegraphics[width=0.5\textwidth]{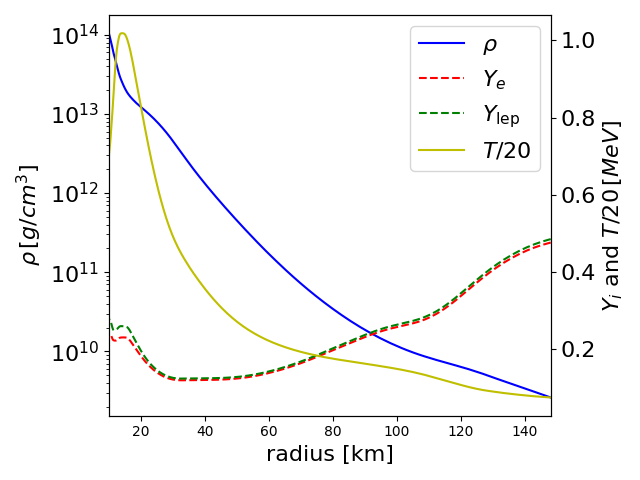}
    \caption{Radial profiles of baryon mass density ($\rho$), electron fraction ($Y_e$), lepton fraction ($Y_{\rm lep}$), and temperature ($T$) for our reference CCSN model. In the plot, color, and line distinguish these thermodynamical quantities. For visualization purposes, we change the scale of $T$ divided by 20.
    }
    \label{fig:combined_plot}
\end{figure}

Before entering into details of CFI analyses, we briefly summarize our CCSN model used as a reference for the present study. Similar to our previous study~\cite{liu2}, we use a fluid profile at the time snapshot of $100\,\mathrm{ms}$ post-bounce in our CCSN model for $15\,M_\odot$ ($M_\odot$ denotes the solar mass) progenitor in \cite{Nagakura_2019a}. In the CCSN simulation, we solved the Boltzmann equation for neutrino transport under multi-energy, multi-angle, and multi-species ($\nu_e$, $\bar{\nu}_e$, and $\nu_x$) treatments. The code simultaneously solves for the hydrodynamics under self-consistent treatments of gravity, hydrodynamics, and neutrino transport. In the simulation, the following neutrino-matter interactions were incorporated: electron-positron pair annihilation, nucleon-nucleon bremsstrahlung, and electron and positron capture by nucleons, heavy and light nuclei as emission processes; their inverse reactions as absorption ones; nucleon scatterings, electron-scatterings, and coherent scatterings with heavy nuclei as scattering processes (see~\cite{Nagakura_2019a} for more details). In Fig.~\ref{fig:combined_plot}, we display radial profiles of representative thermodynamical quantities of the CCSN model: baryon mass density ($\rho$), electron fraction ($Y_e$), lepton number fraction ($Y_{\rm lep}$), and temperature ($T$).

The growth rate of CFI is roughly the same order of magnitude as collision rate (but the resonance-like CFI can have much higher than that of normal one; see below for more details), indicating that the optically thick region is our interest for CFI studies. It should also be noted that on-shell muons can appear only in high-density regions~\cite{PhysRevLett.119.242702,PhysRevD.102.123001}. We, thus, carry out CFI analyses only in the region of the baryon mass density ($\rho$) from $5\times10^{12}\,\mathrm{g/cm^3}$ to $1.0\times10^{14}\,\mathrm{g/cm^3}$. 

\subsection{Fundamental variables and basic assumptions}\label{subsec:funvari}

For our analyses in the following sections, we define some fundamental variables and also describe basic assumptions in this section. We assume that the CCSN matter consists of baryons, charged leptons ($l^{\pm}$), and photons, and they are also assumed to achieve local thermal equilibrium states. The thermodynamical state is determined based on one of the realistic nuclear equations of states (EOS)~\cite{Furusawa_2017}. It should be noted that we extend the EOS to incorporate on-shell muons. Neutrinos, on the other hand, are not in thermal- and chemical equilibrium with matter in general, but the equilibrium state would be almost achieved in optically thick regions which we consider in the present study. It should also be mentioned that the appearance of muons increases opacities of $\barparent{\nu}_\mu$ compared to the case without them. For this reason, we assume that both $\barparent{\nu}_e$ and $\barparent{\nu}_\mu$ reach beta equilibrium with matter. For $\barparent{\nu}_\tau$, we treat them collectively (i.e., they are assumed to be identical). In addition, angular distributions of neutrinos are assumed to be isotropic, but their chemical potentials (or number density) are treated in a parametric manner. As shall be shown later, occurrences of CFI hinge on the chemical potential of $\nu_\tau$. 

The number densities ($n$) and energy densities ($E$) of charged leptons and neutrinos are given as:
\begin{equation}
\begin{split}
    n_i &= \frac{4\pi g_i}{(2\pi\hbar)^3} \int_0^\infty \frac{p^2 \, dp}{e^{(\epsilon_i(p)-\mu_i)/k_B T_i} + 1}, \\
    E_i &= \frac{4\pi g_i}{(2\pi\hbar)^3} \int_0^\infty \frac{p^2 \epsilon_i(p) \, dp}{e^{(\epsilon_i(p)-\mu_i)/k_B T_i} + 1},
\end{split}
\label{nE}
\end{equation}
where the index $i$ represents all charged leptons and neutrinos. Here, $g_i$ is the statistical factor ($=2$ for charged leptons and $=1$ for neutrinos), $\mu_i$ is the chemical potential including rest mass contribution (neutrinos are treated as massless), and $T_i$ is the temperature of the Fermi gas, assumed to be identical for all particles and equal to the baryon temperature in this study. The relation between energy and momentum is given by
\begin{equation}
    \epsilon_i(p) = \sqrt{p^2 c^2 + m_i^2 c^4},
\end{equation}
where $m_i$ is the rest mass of the particle species $i$. The beta equilibrium condition between electrons (muons) and $\barparent{\nu}_e$ ($\barparent{\nu}_\mu$),
\begin{equation}
    e(\mu)^- + p \longleftrightarrow \nu_{e(\mu)} + n,
    \label{eq:betaequ}
\end{equation}
leads to the chemical potential relations:
\begin{equation}
    \mu_{e(\mu)^-} + \mu_p = \mu_{\nu_{e(\mu)}} + \mu_n.
    \label{chemeq1}
\end{equation}

We also note that $e^\pm(\mu^\pm)$ pairs are in equilibrium with photons, leading to the relations among their chemical potentials:
\begin{equation}
    \mu_{e^-} + \mu_{e^+} = \mu_{\mu^-} + \mu_{\mu^+} = 0.
\end{equation}
Together with the chemical potential relations from (inverse) beta equilibrium, this also gives the following relation between the neutrino-anti-neutrino chemical potentials for electron- and muon flavors:
\begin{equation}
    \mu_{\nu_e} + \mu_{\bar{\nu}_e} = \mu_{\nu_\mu} + \mu_{\bar{\nu}_\mu} = 0.
    \label{eq:chemipoteneu_antineu}
\end{equation}

\subsection{Fluid- and neutrino background modeling}\label{subsec:backmod}

As mentioned already, we do not directly use CCSN models with muons in the present study. Instead, we employ fluid distributions taken from CCSN simulations without muons and then add muons by hand in a parametric manner. This approach is straightforward and suited for a systematic study of the impacts of muons on CFI. However, the obtained fluid- and neutrino distributions could significantly deviate from realistic CCSN profiles. This is because the simple addition of muons increases the lepton number and energy, which could be unrealistically higher than those realized in the CCSN core. We, hence, develop a prescription so that both fluid- and neutrino distributions become similar to those typical in the CCSN core.

Given $\rho$, $Y_e$, and $T$ from our CCSN model, we first compute the number- and energy densities of all species of neutrinos assuming that there are no on-shell muons. We can determine their distributions by chemical potentials, which are given by beta equilibrium condition (see Eqs.~\ref{chemeq1}~and~\ref{eq:chemipoteneu_antineu}) for $\nu_e$ and $\bar{\nu}_e$, while they are assumed to be zero for $\barparent{\nu}_\mu$. The chemical potential of $\barparent{\nu}_\tau$ is another parameter in this study, which is set to either $0$, $-0.5$, or $-2\,\mathrm{MeV}$, whose motivation is clarified in section~\ref{subsec:EqBr}. The chosen values are just typical ones obtained from the simulation.

In cases with on-shell muons (i.e., finite $Y_{\mu}$), we change $Y_e$ and $T$ so that the lepton fraction and the total energy density (including neutrinos) become the same as those in the case without on-shell muons. This is a reasonable prescription in very optically thick regions since the lepton number and total energy should not be changed in muon productions at least locally.

There is a caveat, however. In reality, the disparity of weak processes among heavy-leptonic neutrinos causes species-dependent neutrino transport. These advection effects impact lepton number and energy densities quantitatively~\cite{PhysRevD.102.123001}. We should, hence, keep in mind that our prescription is just a technique to suppress unrealistic fluid distributions that largely deviate from those in realistic CCSN environments. A more self-consistent approach is ultimately necessary to draw a robust conclusion for the impacts of on-shell muons on CFI and subsequent CCSN dynamics.

In the present study, we change $Y_\mu$ from 0 to 0.1, which is wide enough to cover the range of interests in CCSN environments. We adopt a uniform grid with 41 points in $Y_{\mu}$ direction, with the grid width of $dY_\mu = 0.0025$. We note that resonance-like CFIs emerge by changing $Y_\mu$ (see Sec.~\ref{sec:results}). Around the resonance region, we increase the $Y_\mu$ resolution by adding 10 additional meshes in each standard $Y_\mu$ grid to capture the resonance feature precisely.

\begin{figure}
    \centering
     \includegraphics[width=0.49\textwidth]{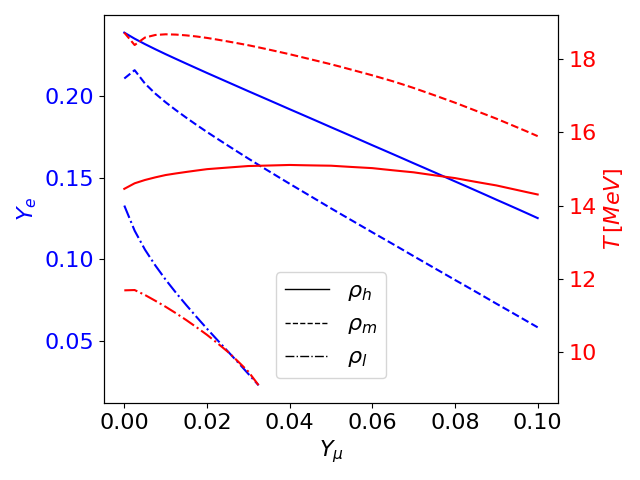}
     \caption{$Y_e$ and $T$ as a function of $Y_\mu$ for selected densities: $\rho_h=1.0\times10^{14}\,\mathrm{g/cm^3}$, $\rho_m=1,5\times10^{13}\,\mathrm{g/cm^3}$, and $\rho_l=7.0\times10^{12}\,\mathrm{g/cm^3}$. The line type distinguishes densities, while blue and red colors represent $Y_e$ and $T$, respectively.}
     \label{them}
\end{figure}

In Fig.~\ref{them}, we display examples of how $Y_e$ and $T$ are changed with $Y_{\mu}$ for three selected baryon mass densities: $\rho_h=1.0\times10^{14}\,\mathrm{g/cm^3}$ (solid line), $\rho_m=1,5\times10^{13}\,\mathrm{g/cm^3}$ (dotted line), and $\rho_l=7.0\times10^{12}\,\mathrm{g/cm^3}$ (dashed line). In the plot, we set the chemical potential of $\barparent{\nu}_\tau$ to be zero, while the response of $Y_e$ and $T$ is less sensitive to it (the difference to the case with $-2\,\mathrm{MeV}$ is about $0.01\%$). As shown in the figure, $Y_e$ in all cases monotonically decreases with $Y_{\mu}$, which is attributed to the conservation of the lepton number. It should be mentioned that the increase of $Y_{\mu}$ also leads to increase (decrease) the number density of $\nu_{\mu}$ ($\bar{\nu}_{\mu}$) due to the beta equilibrium condition, which is complementary by reducing $Y_e$ (and $Y_{\nu_e}-Y_{\bar{\nu}_e}$). In Fig.~\ref{them}, we also find that the response of $T$ depends on $\rho$. For $\rho_h$ (high-density regions), the temperature has a nearly flat profile with respect to changing $Y_{\mu}$. This is because the total energy density is dominated by baryons, and therefore the matter temperature can be less sensitive to $Y_{\mu}$. For lower density regions ($\rho_m$ and $\rho_l$), on the other hand, lepton contributions to the total energy density are non-negligible. As a result, the internal energy of baryons needs to be reduced when we add on-shell muons, leading to the decrease of $T$\footnote{We note that there is a non-monotonic feature of $Y_e$ and $T$ around $Y_{\mu}\sim0$ for $\rho_m$ in Fig.~\ref{them}. This is primarily due to the switch of prescription of $\barparent{\nu}_\mu$. The reference lepton fraction and energy density are taken from a state where $\mu$-type (anti-)neutrinos have 0 chemical potential. Switching to the prescription enforcing beta equilibrium near $Y_\mu=0$ leads to a sudden decrease in $n_{\nu_\mu}-n_{\bar{\nu}_\mu}$. To conserve the lepton fraction, the lepton number in the electron generation must compensate for the decrease in the muon generation.}.

\section{Stability analysis of CFI}
\label{sec:CFI}
The most straightforward way to assess the occurrence of CFI is to solve the dispersion relation obtained by linearizing the quantum kinetic equation (QKE). In our previous study, we derived analytical formulae to quantify the growth rate of CFI~\cite{liu1}. This is a good approximation for isotropic neutrino gases and can significantly reduce computational costs, allowing us to systematically survey occurrences of CFI in CCSN cores~\cite{liu2, CFIakaho}. We do not repeat the derivation here (we refer readers to \cite{liu1,liu2,CFIakaho} for a comprehensive derivation and demonstration), but we summarize these essential formulae.

In our approximate formulae, the CFI growth rates can be given by
\begin{equation}
    \omega_{\pm} = -A - \mathrm{i}\gamma \pm \sqrt{A^2 - \alpha^2 + \mathrm{i}2G\alpha},
    \label{monoew}
\end{equation}
for the isotropy-preserving modes and
\begin{equation}
    \omega_{\pm} = \frac{A}{3} - \mathrm{i}\gamma \pm \sqrt{\left(\frac{A}{3}\right)^2 - \alpha^2 - \mathrm{i}\frac{2}{3}G\alpha},
    \label{monoeisob}
\end{equation}
for the isotropy-breaking ones. In the above equations, the following notations are introduced:
\begin{equation}
    G = \frac{\mathfrak{g} + \bar{\mathfrak{g}}}{2}, \quad A = \frac{\mathfrak{g} - \bar{\mathfrak{g}}}{2}, \quad \gamma = \frac{\Gamma + \bar{\Gamma}}{2}, \quad \alpha = \frac{\Gamma - \bar{\Gamma}}{2},
    \label{defGAgmal}
\end{equation}
where
\begin{equation}
    \mathfrak{g} = \sqrt{2}G_\text{F}(\hbar c)^2 \left(n_{\nu_i} - n_{\nu_j}\right),
    \label{eq:gdef}
\end{equation}
and 
\begin{equation}
    \Gamma = \frac{\Gamma_{\nu_i} + \Gamma_{\nu_j}}{2}.
\end{equation}
In the expression, $G_F$, $\hbar$, and $c$ denote the Fermi constant, the reduced Planck constant, and the speed of light, respectively; $i$ and $j$ represent neutrino flavors. It should be mentioned that there are three different sectors for potential occurrences of CFI: $e\mu$, $e\tau$, and $\mu\tau$. We use the same convention for anti-neutrinos, while their quantities are highlighted by barred indices.

The number density of neutrinos and mean collision rates are computed by integrating over momentum space:
\begin{equation}
\begin{split}
    n_{\nu_i} &= \frac{1}{(2\pi\hbar)^3} \int dP \, f_{\nu_i}(P), \\
    \Gamma_{\nu_i} &= \langle\Gamma\rangle_{\nu_i} = \frac{1}{n_{\nu_i}} \frac{1}{(2\pi\hbar)^3} \int dP \, \Gamma_{\nu_i}(P) f_{\nu_i}(P),
\end{split}
\label{nGM}
\end{equation}
where $\int dP = 4\pi \int_0^\infty p^2 dp$ ($p$ denotes the absolute value of neutrino momentum).
$f_{\nu_i}$
corresponds to neutrino distributions, which are assumed to be Fermi-Dirac distributions with a chemical potential ($\mu_{\nu}$) and temperature ($T$), indicating that the resultant neutrino number density is identical to Eq.~\ref{nE}. $\Gamma_{\nu_i}(P)$ denotes the collision rate for neutrino flavor of $i$, whose detailed information is described in appendix. We note that the quantities of $G$, $A$, $\gamma$, and $\alpha$ can be written in the same unit of the growth rate of flavor conversions.

In the high-density region, the inequality $\mathfrak{g}_{\nu_i}\gg\Gamma_{\nu_i}$ is usually satisfied. Assuming the inequality, Eqs.~\ref{monoew} and \ref{monoeisob} can be rewritten in a more concise form,
\begin{equation}
\text{max\,\lbrack Im\,}\omega\rbrack=\begin{cases}
    -\gamma+\frac{|G\alpha|}{|A|},& \text{if }A^2\gg |G\alpha|,\\
    -\gamma+\sqrt{|G\alpha|},& \text{if }A^2\ll |G\alpha|,
\end{cases}
\label{wapprpre}
\end{equation}
for the isotropy-preserving branch and
\begin{equation}
\text{max\,\lbrack Im\,}\omega\rbrack=\begin{cases}
    -\gamma+\frac{|G\alpha|}{|A|},& \text{if }A^2\gg |G\alpha|,\\
    -\gamma+\frac{\sqrt{|G\alpha|}}{\sqrt{3}},& \text{if }A^2\ll |G\alpha|,
\end{cases}
\label{wapprbr}
\end{equation}
for the isotropy-breaking one. In the two sets of equations, the first lines correspond to the non-resonance case and the second lines correspond to the resonance-like enhancement one for the isotropy-preserving and isotropy-breaking branches, respectively. As a matter of fact, it has been shown in \cite{liu1} that the condition for the maximum growth rates occurs at $|A|=|\alpha|$ and $|A|=|3\alpha|$ (the signs depend on the signs of $G$ and $\alpha$) for the isotropy-preserving and isotropy-breaking branches, respectively. One thing we should mention is that the growth rate of CFIs is computed based on Eqs.~\ref{monoew}~and~\ref{monoeisob}, while other approximate formulae are used to interpret the properties of CFIs.

\begin{figure*}
    \centering
    \subfigure[]{\includegraphics[width=0.49\textwidth]{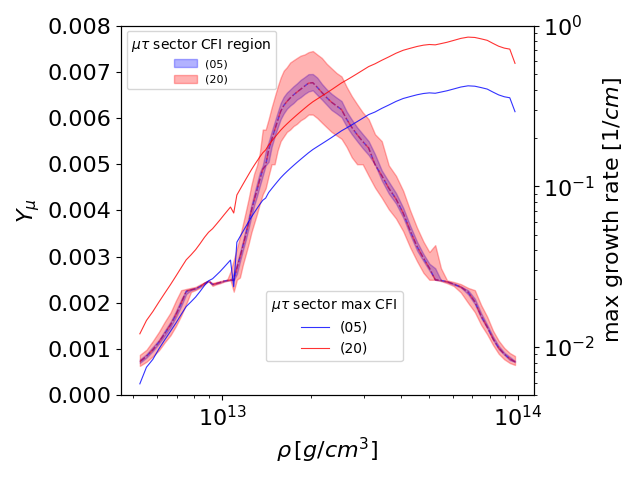}
         \label{mutaubound}
     }
     \subfigure[]{\includegraphics[width=0.49\textwidth]{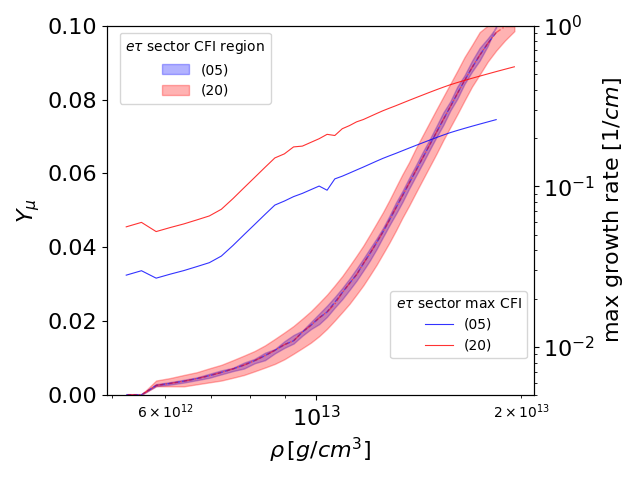}
         \label{etaubound}
     }
     \caption{
     $Y_\mu$ ranges for CFI (shaded region) and their maximum growth rates (solid lines) as a function of $\rho$. The left and right axes in each panel represent $Y_{\mu}$ and CFI growth rate, respectively. The left panel is for the $\mu\tau$ sector, and the right panel is for the $e\tau$ sector. Different colors correspond to the different cases with chemical potential of $\barparent{\nu}_{\tau}$: $\mu_{\nu_\tau}=\mu_{\bar{\nu}_\tau}=-0.5\,\mathrm{MeV}$ (blue) and $-2.0\,\mathrm{MeV}$ (red).
     }
     \label{CFIbound}
\end{figure*}

\section{Results}
\label{sec:results}
In this section, we present and discuss the results of our analyses for CFI with on-shell muons. It is important to mention that we find no unstable solutions of CFI in the $e\mu$ sector. We also find that in the case with $\mu_{\barparent{\nu}_{\tau}}=0$, no CFIs occur in all three neutrino-mixing sectors. In Sec.~\ref{subsec:EqBr}, we first prove these results analytically and then make a statement that breaking the beta (pair) equilibrium for $\barparent{\nu}_e$, $\barparent{\nu}_\mu$, or $\barparent{\nu}_\tau$ is a necessary condition to trigger CFI. In the rest of the subsections, we delve into our results in cases with $\mu_{\barparent{\nu}_{\tau}} \neq 0$.

\subsection{No CFIs in beta- and pairs equilibrium states} \label{subsec:EqBr}

First, let us focus on CFI in the $e\mu$ sector. As described in Sec.~\ref{subsec:funvari}, $\barparent{\nu}_e$ and $\barparent{\nu}_\mu$ are assumed to be in beta equilibrium with identical temperatures. The chemical potential of each species of neutrinos, hence, satisfies the relation of $\mu_{\nu_i}=-\mu_{\bar{\nu}_i}$, where $i$ denotes $e$ and $\mu$. We note that $i$-flavor neutrino lepton number, $\left(n_{\nu_i}-\bar{n}_{\nu_i}\right)$, is a monotonic function of $\mu_{\nu_i}$ under the same temperature, which leads to the following relation,
\begin{equation}
    \mu_{\nu_i} \geq  \mu_{\nu_j} \Longleftrightarrow (n_{\nu_i} - n_{\bar{\nu}_i}) \geq (n_{\nu_j} - n_{\bar{\nu}_j}).
    \label{eq:chemi_lead_leptonum}
\end{equation}

On the other hand, $A$ is proportional to the difference of neutrino lepton numbers between $i$ and $j$ flavors (see also Eqs.~\ref{defGAgmal}~and~\ref{eq:gdef}), i.e.,
\begin{equation}
    A\propto\left[\left(n_{\nu_i}-\bar{n}_{\nu_i}\right)-\left(n_{\nu_j}-\bar{n}_{\nu_j}\right)\right].
    \label{eq:Apropdiflepnum}
\end{equation}
From Eqs.~\ref{eq:chemi_lead_leptonum}~and~\ref{eq:Apropdiflepnum}, $A=0$ can achieve only if $\mu_{\nu_i} = \mu_{\nu_j}$, which also leads to the identical neutrino number densities between $i$ and $j$ flavors. This implies $\mathfrak{g}=\bar{\mathfrak{g}}=0$, and consequently $G$ also becomes zero (see also Eq.~\ref{defGAgmal}). From Eqs.~\ref{wapprpre}~and~\ref{wapprbr}, the condition of $G=0$ always leads to negative $\text{Im\,}\omega$, indicating that CFIs do not occur.

Next, let's consider cases with $A>0$, while we omit the discussion with $A<0$ since the same conclusion can be obtained by exchanging flavors between $i$ and $j$ from the case with $A>0$. From Eqs.~\ref{eq:chemi_lead_leptonum}~and~\ref{eq:Apropdiflepnum}, $A>0$ leads to $ \mu_{\nu_i} >  \mu_{\nu_j}$ and $\mu_{\bar{\nu}_i} <  \mu_{\bar{\nu}_j}$. As a result, we can obtain $\mathfrak{g}>0$, and $\bar{\mathfrak{g}}<0$ (see also Eq.~\ref{eq:gdef}), which guarantees the following inequality relation,
\begin{equation}
    \left|\mathfrak{g} + \bar{\mathfrak{g}}\right| < |\mathfrak{g}| + |\bar{\mathfrak{g}}| = |\mathfrak{g} - \bar{\mathfrak{g}}|.
    \label{eq:ineq_g_Alargezero}
\end{equation}
From the definitions of $G$ and $A$ (see Eq.~\ref{defGAgmal}) and Eq.~\ref{eq:ineq_g_Alargezero}, we can obtain $|G|<|A|$, which corresponds to a key relation to see no occurrences of CFI. For cases with $A^2\gg |G\alpha|$ in Eqs.~\ref{wapprpre}~and~\ref{wapprbr}, the condition of $|G|<|A|$ results in negative $\text{Im\,}\omega$ due to $|\alpha| < |\gamma|$. For cases with $A^2\ll |G\alpha|$, on the other hand, the condition of $|\alpha| \gg |A|$ should be satisfied due to $|G|<|A|$, which leads to $\sqrt{|G \alpha|} < \sqrt{|\alpha^2|} < |\gamma|$. As a result, $\text{Im\,}\omega$ is always negative, i.e., no occurrences of CFIs.

The similar argument can also be made for $e \tau$ and $\mu \tau$ sectors in the case with $\mu_{\nu_{\tau}} = \mu_{\bar{\nu}_{\tau}} = 0$, i.e., the pair equilibrium for $\barparent{\nu}_\tau$. As discussed above, the condition of $|G|<|A|$ always satisfies in each sector, indicating that CFI is forbidden. This conclusion is consistent with our finding in our previous studies~\cite{liu2,CFIakaho}, in which we found no occurrences of CFIs in very optically thick regions where heavy-leptonic neutrinos are nearly in pair equilibrium.

From the above argument, CFIs can occur in $e\tau$ and $\mu\tau$ sectors at regions where $\barparent{\nu}_{\tau}$ are out of pair equilibrium. One thing we do notice here is that $\barparent{\nu}_{\tau}$ decouples with matter inside the energy spheres of $\barparent{\nu}_{e}$ and $\barparent{\nu}_{\mu}$ due to the lack of charged current reactions, while their angular distributions can remain isotropic by large opacities of nucleon scatterings. This region is our interest in the present study. In the following sections, we discuss these types of CFIs in more detail.

\subsection{Overall property}
\label{subsec:CFIrho}
One of the striking results of the present study is that CFIs can be induced by the appearance of muons in regions where they are stable in cases without on-shell muons. As shown in our previous paper \cite{liu2}, regions with the baryon mass density of $\rho \gtrsim 10^{13} {\rm g/cm^3}$ are stable with respect to CFI in cases without muons\footnote{But multi-dimensional effects such as PNS convection can offer environments for resonance-like CFIs in the region; see \cite{CFIakaho} for more details}. This result is consistent with our discussion in the previous section that all species of neutrinos are nearly in thermal equilibrium with matter, which hampers CFIs. As shown below, however, the appearance of on-shell muons can induce CFIs even in such very optically thick regions, and more importantly, the instability condition is met even by small deviations of $\barparent{\nu}_{\tau}$ from the pair equilibrium. Understanding these properties of CFIs is the main focus of this subsection.

Figure~\ref{CFIbound} displays the summary of our results. As shown in the figure, CFIs can occur in certain $Y_{\mu}$ ranges for both $\mu\tau$ (left panel) and $e\tau$ (right panel) sectors. As shown in the figure, the growth rate is higher, and the unstable region of $Y_{\mu}$ is wider for the lower chemical potential of $\barparent{\nu}_{\tau}$. These trends are consistent with our claim that the degree of breaking equilibrium accounts for occurrences of CFIs.

We also find that resonance-like CFIs are involved in both sectors. In fact, the maximum growth rate at each $\rho$ in Fig.~\ref{CFIbound} is given at the resonance point. The $Y_{\mu}$ at the resonance point (denoted as $Y_{\mu,c}$ hereafter), is insensitive to $\mu_{\barparent{\nu}_{\tau}}$. This feature is displayed in Fig.~\ref{CFIbound}, in which the dashed lines of red and blue colors are overlapped with each other. The trend can be understood as follows. For $\mu \tau$ sector, the resonance condition is given by $n_{\nu_\mu} \sim n_{\bar{\nu}_\mu}$, i.e., $\mu_{\nu_\mu} \sim 0$. This implies that the chemical potential of muons ($\mu_{\mu,c}$) at the resonance point (or $Y_{\mu, c}$) can be determined as (see also Eq.~\ref{chemeq1})
\begin{equation}
    \mu_{\mu,c} \sim \mu_n-\mu_p.
    \label{eq:critmu}
\end{equation}
Since both $\mu_n$ and $\mu_p$ are insensitive to $\nu_{\tau}$\footnote{Strictly speaking, chemical potentials of baryons depend on $\mu_{\barparent{\nu}_{\tau}}$ since all thermodynamical states are determined for each $Y_{\mu}$ so that the total energy density is conserved. However, the contribution of $\barparent{\nu}_{\tau}$ is always subdominant, and therefore is negligible.}, Eq.~\ref{eq:critmu} implies that $\mu_{\mu,c}$ also does not depend on $\nu_{\tau}$.

A similar argument can be applied to the $e \tau$ sector. The critical chemical potential of the electron, that meets a condition of resonance-like CFI, can be given as
\begin{equation}
    \mu_{e,c} \sim \mu_n-\mu_p.
    \label{eq:crite}
\end{equation}
One thing we do notice here is that $\mu_e$ is usually higher than $\mu_n-\mu_p$ in CCSN environments, which is observed in the entire domain for our spherically symmetric CCSN model. This indicates that CFIs in the $e \tau$ sector can be triggered by reducing $\mu_e$ (or $Y_e$). As discussed in Sec.~\ref{subsec:backmod} (see also Fig.~\ref{them}), the appearance of muons leads to the reduction of $Y_e$ to conserve the lepton number, causing occurrences of CFIs in $e \tau$ sector.

\begin{figure}
    \centering
    \includegraphics[width=0.49\textwidth]{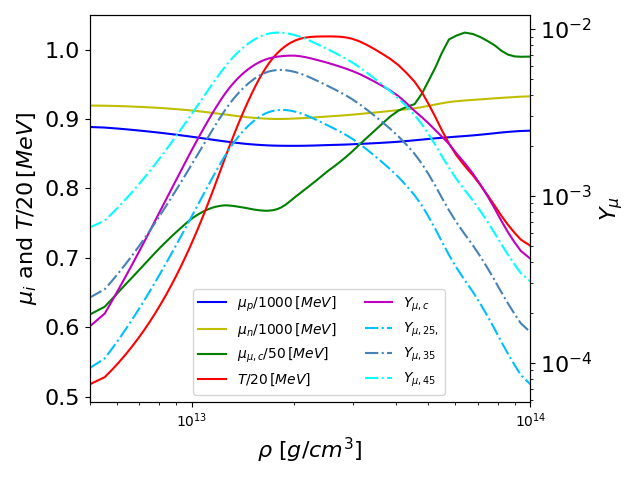}
     \caption{Radial profiles of some key quantities to understand the non-monotonic trend of $Y_{\mu, c}$ in $\mu \tau$ sector (see the left panel of Fig.~\ref{CFIbound}). We plot $\mu_p$ (blue), $\mu_n$ (gold), $\mu_{\mu, c}$ (green), $T$  (red), and $Y_{\mu, c}$ (purple) by solid lines. We also provide three different $Y_{\mu}$, in which the muon fraction is computed under the assumption that $\mu_{\mu}$ is fixed ($25$, $35$, and $45\,\mathrm{MeV}$) but adopting temperature profile from the original one. See text for more details.}
     \label{fig:critmu}
\end{figure}

By comparing the left and right panels in Fig.~\ref{CFIbound}, we also find that the $Y_{\mu}$ distributions for CFIs as a function of $\rho$ are very different between $\mu \tau$ and $e \tau$ sectors. In the $e \tau$ sector (right panel), it is roughly a monotonic function of $\rho$, whereas it has a peak profile for the $\mu \tau$ sector (left panel). Let us first consider the case in the $\mu \tau$ sector. To this end, we provide radial profiles of some key quantities in Fig.~\ref{fig:critmu}. One thing we do notice here is that temperature has a similar peak profile (see the red line in Fig.~\ref{fig:critmu}), whereas $\mu_{\mu, c}$ roughly increases with $\rho$. This indicates that temperature distribution accounts for the non-monotonic profile of $Y_{\mu}$. To verify our hypothesis, we compute other three $Y_{\mu}$ profiles, in which we compute the muon fraction by employing the same temperature profile but assuming a constant $\mu_{\mu}$: $25$, $35$, and $45$ MeV. As shown in Fig.~\ref{fig:critmu}, obtained $Y_{\mu}$ profiles have a similar trend as $Y_{\mu, c}$ (red solid line). We also note that the decrease of temperature in the high baryon density directly reduces $n_{\mu} - n_{\bar{\mu}}$, which makes $Y_{\mu, c} \propto (n_{\mu} - n_{\bar{\mu}})/\rho$ a sharp decline with $\rho$.

\begin{figure}
    \centering
     \includegraphics[width=0.49\textwidth]{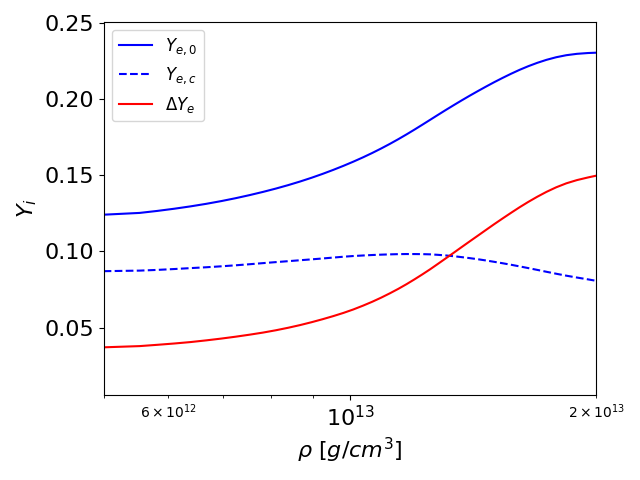}
     \caption{$Y_i$ as a function of $\rho$ for analysis of CFI in $e \tau$ sector. Here, $Y_{e,0}$ denotes electron fraction at $Y_\mu=0$; $Y_{e,c}$ represents a critical electron fraction to trigger resonance-like CFIs, and $\Delta Y_e$ is defined as $Y_{e,0}-Y_{e,c}$.}
     \label{fig:yec}
\end{figure}

For the $e\tau$ sector (see in the right panel of Fig.~\ref{CFIbound}), on the other hand, $Y_{\mu}$ leading to unstable CFI tends to increase with $\rho$. Below, we analyze such a positive correlation. In Fig.~\ref{fig:yec}, we display the electron fraction in the case with $Y_{\mu}=0$ (solid blue line) denoted by $Y_{e, 0}$, that computed with the critical chemical potential of electrons (see Eq.~\ref{eq:crite}) denoted by $Y_{e, c}$ (dashed blue line), and their difference $\Delta Y_e$ (solid red line). It should be mentioned that $\Delta Y_e$ corresponds to the required reduction of $Y_e$ for occurrences of resonance-like CFI, indicating that it is nearly equal to $Y_{\mu, c}$ in $e\tau$ sector (see right panel of Fig.~\ref{CFIbound}). As can be seen clearly, the monotonic increase of $\Delta Y_e$ traces the same trend as in $Y_{e, 0}$. The increase of $Y_{e, 0}$ with $\rho$ reflects the fact that neutrinos are trapped in the high-density region, implying that deleptonization is suppressed.

There is an important remark to be made here. Although we show
that CFIs in $e \tau$ sector is limited in the region of $\rho \lesssim 2 \times 10^{13} {\rm g/cm^3}$, they
could occur in higher-density regions in the late phase. This is because the deleptonization of the CCSN core proceeds with time, indicating that the required amount of $Y_{\mu}$ for occurrences CFI would also be reduced. In addition to this, PNS convection, which is neglected in our spherically symmetric CCSN model, would also further accelerate the deleptonization of PNS core~\cite{10.1093/mnras/staa261}, which would offer preferable environments for occurrences of CFIs. It should also be mentioned that, during the accretion phase of CCSN, matter temperature in the PNS envelope increases with time by PNS contraction. This suggests that the average energy of $\barparent{\nu}_{\mu}$ increases with time, which would promote the production of
on-shell muons~\cite{PhysRevD.102.123001}. This consideration motivates a more systematic study of CFIs with muons by covering many different time snapshots under appropriate treatments of multi-dimensional fluid instabilities. We defer this detailed study to future work.

\begin{figure*}
    \centering
    \subfigure[]{\includegraphics[width=0.49\textwidth]{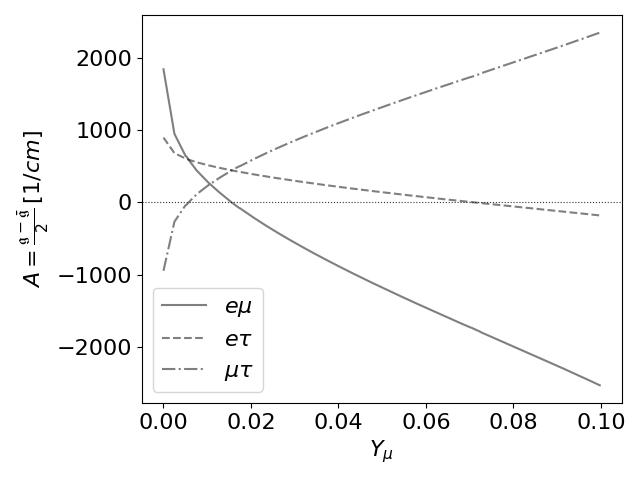}
         \label{A}
     }
     \subfigure[]{\includegraphics[width=0.49\textwidth]{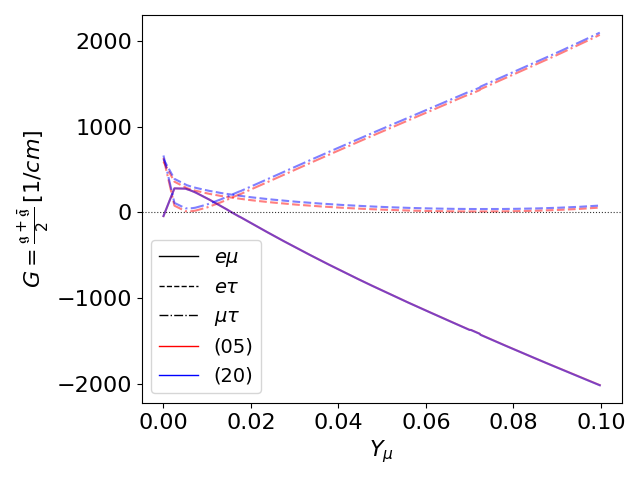}
         \label{G}
     }
     \subfigure[]{\includegraphics[width=0.49\textwidth]{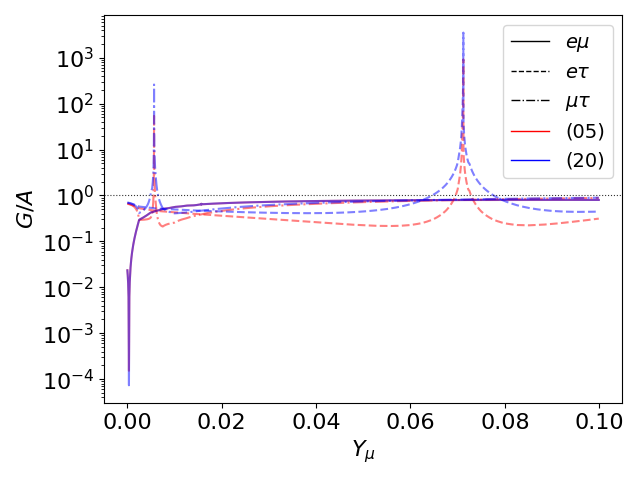}
         \label{GA}
     }
     \subfigure[]{\includegraphics[width=0.49\textwidth]{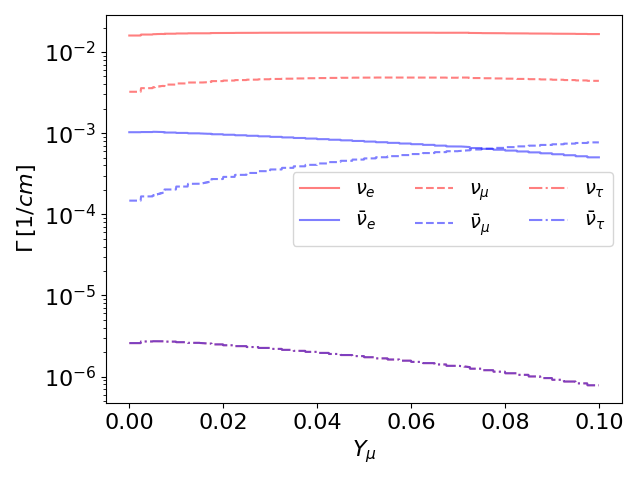}
         \label{GM}
     }
     \caption{Key quantities to assess CFIs, displayed as a function of $Y_{\mu}$. We show $A$, $G$, $G/A$, and $\Gamma$ in the panels of (a), (b), (c), and (d), respectively. The line type distinguishes mixing sectors. Red and blue colors represent the case with $\mu_{\tau}=\mu_{\bar{\tau}}=-0.5$MeV and $-2$MeV, respectively.
     }
     \label{AGGM}
\end{figure*}

\subsection{Local analysis}
\label{subsec:onepoint}
In this subsection, we inspect more closely how CFIs can be induced by the appearance of on-shell muons. As a representative example, we focus upon a spatial mesh point where the baryon mass density, electron fraction, and temperature (in the state without on-shell muons) are $\rho=1.5\times10^{13}\,\mathrm{g/cm^3}$, $Y_e = 0.21$, and $T = 18.7$ MeV, respectively. We note that CFIs can be unstable in both $e \tau$ and $\mu \tau$ sectors in a reasonable range of $Y_{\mu}$ (see Fig.~\ref{CFIbound}).

In Fig.~\ref{AGGM}, we display how key variables ($G$, $A$, $\Gamma$, and their combinations) are changed with $Y_{\mu}$ at all neutrino-mixing sectors. It should be mentioned that CFIs do not occur in $e \mu$, as discussed in Sec.~\ref{subsec:EqBr}, although the result in the sector is also shown as a reference. In the top panel of the figure, we portray $A$ as a function of $Y_{\mu}$. As shown in the panel, $A$ in $e \mu$ and $e \tau$ decreases with $Y_{\mu}$. This is because the increase of $Y_{\mu}$ reduces $Y_e$ to conserve the lepton number. As a result, $n_{\nu_e}$ ($n_{\bar{\nu}_e}$) also decreases (increases) with $Y_{\mu}$, leading to $dA/d{Y_{\mu}} <0$. In $\mu \tau$ sector, on the other hand, $A$ increases with $Y_{\mu}$, which is simply because $\nu_{\mu}$ ($\bar{\nu}_{\mu}$) monotonically increases (decreases) with $Y_{\mu}$.

We also find that $A$ changes the sign at a certain value of $Y_{\mu}$, indicating that resonance-like CFIs potentially occur. As shown in the top right panel, however, $G$ also becomes zero at the $Y_{\mu}$ satisfying $A=0$ in the $e \mu$ sector, implying no occurrences of CFIs. On the other hand, $G$ remains finite for other sectors, indicating that the resonance-like CFI can occur in these sectors. Such features can be observed in the bottom left panel, in which $G/A$ has diverse features in $e \tau$ and $\mu \tau$ sectors. As shown in the plot, the resonance-like CFI in the $e\tau$ sector occurs at much higher $Y_\mu$ than that of the $\mu\tau$ one, which is consistent with our discussion in Sec.~\ref{subsec:CFIrho}.

We also provide reaction rates of neutrino-matter interactions for each species of neutrinos in the bottom right panel of Fig.~\ref{AGGM}. As shown in the panel, those for $\barparent{\nu}_\tau$ are negligibly low. On the other hand, $\barparent{\nu}_{\mu}$ have much higher rates than $\barparent{\nu}_\tau$. It is worth noting that the maximum growth rate of CFI in the $\mu \tau$ sector is higher than that in $e \tau$ one (see Fig.~\ref{CFIbound}). One might think that the trend is inconsistent with the trend of $\Gamma$ in Fig.~\ref{AGGM} since $\barparent{\nu}_{\mu}$ reactions are nearly one order of magnitude lower than those for $\barparent{\nu}_{e}$. One thing we do notice here is that the growth rate of resonance-like CFI is roughly proportional to $\sqrt{G \alpha}$ (see Eqs.~\ref{wapprpre}~\ref{wapprbr}), indicating that the growth rate can be higher by large $G$ even if $\alpha$ is relatively small. In fact, the CFI in $\mu \tau$ sector can occur in a higher density than that in $e \tau$ sector (see the density region in Fig.~\ref{CFIbound}\footnote{As we have already pointed out, the CFI condition in $e \tau$ sector can be met if $Y_{\mu}>0.1$ in the high-density region, but such extreme high $Y_{\mu}$ environment would not be realized in realistic CCSN core.}). This accounts for the disparity of growth rate between $\mu \tau$ and $e \tau$ sectors.

\begin{figure*}
    \centering
    \subfigure[]{\includegraphics[width=0.49\textwidth]{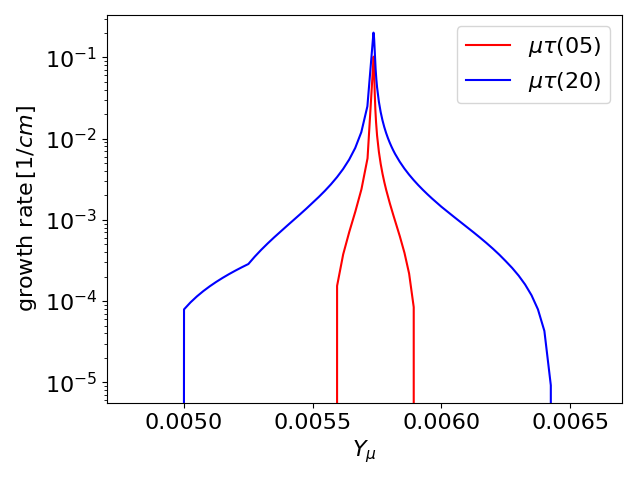}
         \label{CFI1}
     }
     \subfigure[]{\includegraphics[width=0.49\textwidth]{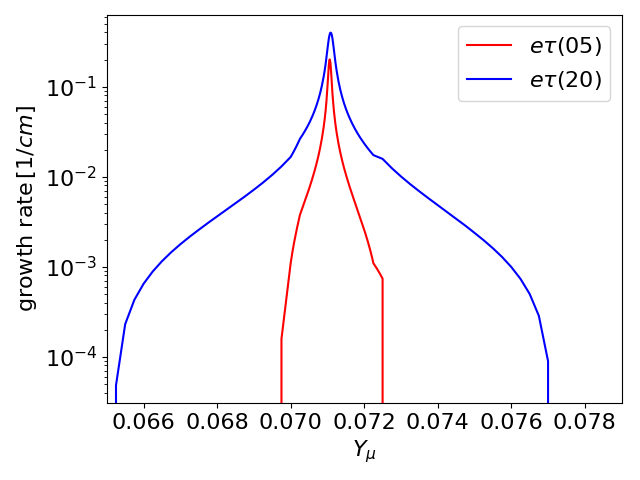}
         \label{CFI2}
     }
     \caption{CFI growth rate for the $\mu\tau$ sector (left panel) and the $e\tau$ sector (right panel) as a function of $Y_{\mu}$. Red and blue colors represent the case with $\mu_{\tau}=\mu_{\bar{\tau}}=-0.5$MeV and $-2$MeV, respectively.
     }
     \label{1p}
\end{figure*}

In Fig.~\ref{1p}, we display growth rates of CFIs as a function of $Y_{\mu}$. As shown in the figure, they have peak profiles, and they reach the maximum at $Y_\mu \sim 0.0057$ in the $\mu\tau$ sector and at $Y_\mu \sim 0.071$ in the $e\tau$ sector. We also find that $Y_{\mu}$ of resonance point in both sectors do not depend on $\mu_{\bar{\nu}_{\tau}}$, whereas there are remarkable differences in the overall growth rate profile between the two different $\mu_{\bar{\nu}_{\tau}}$ cases. In the case with $\mu_{\bar{\nu}_{\tau}}=-2$MeV, the range of $Y_{\mu}$ with CFIs becomes wider and the maximum value of the growth rate is higher than those in the case with $\mu_{\bar{\nu}_{\tau}}=-0.5$MeV. These results also agree with the discussion of Fig.~\ref{AGGM}.

\subsection{Impacts of CFIs with muons on CCSN}\label{subsec:impacts}

Finally, let us discuss the possible impacts of CFIs with muons on neutrino radiation fields. We should keep in mind, however, that the discussion is highly provisional, and more self-consistent studies including solving non-linear QKE are needed to validate the following discussion. 

When CFIs occur in $i \tau$ sector ($i$ denotes $e$ or $\mu$), flavor conversion mixes up $\nu_{i} (\bar{\nu}_{i})$ and $\nu_{\tau} (\bar{\nu}_{\tau})$ or totally swaps them in the case of resonance~\cite{PhysRevD.109.103009}. On the other hand, CFIs cannot convert neutrinos to anti-neutrinos or vise-versa, implying that there remains a disparity between neutrinos and anti-neutrinos in the non-linear phase. This implies that $\nu_{\tau}$ and $\bar{\nu}_{\tau}$ can be no longer treated collectively. Although it has already been suggested that the weak magnetism can differentiate $\nu_{\tau}$ and $\bar{\nu}_{\tau}$ in CCSNe~\cite{PhysRevLett.119.242702}, CFIs might enhance the difference.

One of the important aspects of CFIs induced by muons is that these flavor instabilities can occur in very optically thick regions where $\barparent{\nu}_e$ and $\barparent{\nu}_\mu$ are nearly in thermal- and chemical equilibrium with matter (but $\barparent{\nu}_\tau$ needs to deviate from the pair equilibrium at least to a certain extent). If flavor conversions occur in such high-density region ($\rho \gtrsim 10^{13} {\rm g/cm^3}$), they would have enormous impacts on CCSN dynamics and neutrino signals~\cite{PhysRevD.107.103034,PhysRevLett.131.061401}. On the other hand, all previous work of CFIs (and also FFC) do not distinguish $\nu_{\mu}$ and $\nu_{\tau}$, exhibiting that the non-linear evolution of flavor conversions in cases with on-shell muons would be qualitatively different from that in the case without them. In fact, $e \tau$ and $e \mu$ sectors are no longer identical, indicating that complex interplay among all three mixing sectors would emerge. Addressing this issue is very intriguing and deserves further investigation, which we leave to our future work.

\section{Conclusion}\label{sec:conclusion}
In this paper, we investigated the stability properties of CFIs with on-shell muons in CCSNe, motivated by theoretical indications that on-shell muons could be produced in the PNS envelope in the post-bounce phase. The striking results of the present study are summarized as follows.

We first analytically show that breaking of beta- or pair equilibrium in each neutrino species is a necessary condition to trigger CFIs. In CCSN environments, $\barparent{\nu}_{\tau}$ represents the most transparent neutrino species due to the lack of charged-current reactions, indicating that CFIs in $e \tau$ and $\mu \tau$ sectors could occur in the high-density region. In fact, we provide evidence that they can occur in the density region of $10^{12} {\rm g/cm^3} \lesssim \rho \lesssim 10^{14} {\rm g/cm^3}$. We also find that the required $Y_{\mu}$ to induce CFIs in $\mu \tau$ sector is, in general, lower than that in $e \tau$, which allows occurrences of CFIs in $\mu \tau$ sector at the higher density region ($\rho \gg 10^{13} {\rm g/cm^3}$).

Another important finding in the present study is that resonance-like CFIs can occur in $\mu \tau$ ($e \tau$) sector when $\mu_{\mu}$ ($\mu_{e}$) equals to $\mu_{n} - \mu_{p}$. It is worth noting that no explicit dependence on muons is involved in the $e \tau$ sector, but the appearance of muons accounts for the CFI in the sector. This is because the appearance of muons reduces $Y_e$, implying the change in the chemical potential of electrons. As a result, the disparity of number densities between $\nu_e$ and $\bar{\nu}_e$ is reduced in CCSN environments, which can trigger CFIs in the $e \tau$ sector. 

The growth rate of resonance-like CFIs can be much larger than neutrino-matter interactions, exhibiting a possibility of large impacts of CFIs on CCSN dynamics. In fact, charged-current reaction rates with muons are lower than those with electrons, but we show that the growth rate exceeds $0.1 {\rm cm^{-1}}$ (see Fig.~\ref{CFIbound}). This is comparable to or even could be higher than FFC, exhibiting a large potential to impact CCSN.

The present study stimulates further interest in the details of CFIs with muons. Although qualitative trends presented in this paper would not be changed, there are still many details to be worked out before drawing any conclusions. More self-consistent treatments of input physics in CCSN modeling are certainly necessary, for instance, incorporating the effects of multi-dimensional fluid instabilities and weak interactions with muons. 

Our results also underscore the importance of integrating heavy leptons and fully resolving all neutrino flavors in CCSN modeling. Addressing this issue requires appropriate treatments of neutrino quantum kinetics coupled with other relevant physical processes, which will eventually shed light on flavor conversion mechanisms, their impact on the explosion dynamics, the nucleosynthesis thereafter, and the observable neutrino spectra. This is obviously not an easy task, but we will tackle the issue in collaboration with the CCSN- and neutrino-oscillation communities.

Last but not least, a final remark regarding binary neutron star merger (BNSM) is provided here. As shown in previous studies (see, e.g., \cite{muonBNS}), on-shell muons could appear in the post-merger phase. The present study can certainly be an important reference for the study, but both fluid- and neutrino fields are remarkably different between CCSNe and BNSMs. Detailed investigations of CFIs in BNSM environments are, hence, necessary, which is also on our to-do list. The results will be reported in forthcoming papers.

\begin{acknowledgments}
We thank Ken’ichi Sugiura for helping calculate weak interaction rates. 
Numerical computation in this work was partly carried out at the Yukawa Institute Computer Facility.
This work is also supported by the HPCI System Research Project (Project ID:240079).
H.N. is supported by Grant-in-Aid for Scientific Research (23K03468).
M.Z. is supported by a JSPS Grant-in-Aid for JSPS Fellows (Grant No. 22KJ2906) from the Ministry of Education, Culture, Sports, Science, and Technology (MEXT) in Japan.
S.F. is supported by Grant-in-Aid for Scientific Research (19K14723 and 24K00632).
S.Y. is supported by Grant-in-Aid for Scientific Research (21H01083).
S.Y. is supported by the Institute for Advanced Theoretical and Experimental Physics,
Waseda University, and the Waseda University Grant for Special Research Projects (project No. 2023C-141, 2024C-56, 2024Q-014).
\end{acknowledgments}
\bibliographystyle{apsrev4-2}
\bibliography{sample}

\appendix
\section{Neutrino Interactions}
We evaluate the following neutrino-matter interactions in this study. The following six groups of interactions contribute to CFI in this study.\\
The semi-leptonic interactions:
\begin{equation}
    \begin{split}
        \text{electron capture}\quad e^-+p\longleftrightarrow n+\nu_e \\
        \text{positron capture}\quad e^++n\longleftrightarrow p+\bar{\nu}_e\\
        \text{muon capture}\quad\mu^-+p\longleftrightarrow n+\nu_\mu\\
        \text{anti-muon capture}\quad\mu^++n\longleftrightarrow p+\bar{\nu}_\mu
    \end{split}
    \label{ccr}
\end{equation}
The neutrino-lepton flavor exchange interactions:
\begin{equation}
    \begin{split}
        \quad\mu^- + \bar{\nu}_\mu \longleftrightarrow  e^- + \bar{\nu}_e\\
        \quad\mu^- + \nu_e \longleftrightarrow  e^- + \nu_\mu\\
        \quad\mu^+ + \nu_\mu \longleftrightarrow  e^+ + \nu_e\\
        \quad\mu^+ + \bar{\nu}_e \longleftrightarrow  e^+ + \bar{\nu}_\mu
    \end{split}
    \label{reaction:FE}
\end{equation}
The leptonic $e\mu$ annihilation interactions:
\begin{equation}
    \begin{split}
        \quad e^- + \mu^+ \longleftrightarrow  \nu_e + \bar{\nu}_\mu\\
        \quad e^+ + \mu^- \longleftrightarrow  \bar{\nu}_e + \nu_\mu
    \end{split}
\end{equation}
The muon decay interactions:
\begin{equation}
    \begin{split}
        \text{muon decay}\quad\mu^-\longleftrightarrow e^-+\bar{\nu}_e+\nu_\mu\\
        \text{anti-muon decay}\quad\mu^+\longleftrightarrow e^++\nu_e+\bar{\nu}_\mu
    \end{split}
\end{equation}
The nucleon decay interactions:
\begin{equation}
    \begin{split}
        \text{neutron decay}\quad\bar{\nu}_e + p + e^- \longleftrightarrow  n\\
        \text{neutron decay}\quad\bar{\nu}_\mu + p + \mu^- \longleftrightarrow  n\\
        \text{positron emission}\quad\nu_e + n + e^+ \longleftrightarrow  p\\
        \text{anti-muon emission}\quad\nu_\mu + n + \mu^+ \longleftrightarrow  p
    \end{split}
    \label{nd}
\end{equation}
The lepton pair annihilation interactions:
\begin{equation}
    \begin{split}
        \quad e^- + e^+ \longleftrightarrow  \nu + \bar{\nu}\\
        \quad\mu^- + \mu^+ \longleftrightarrow  \nu + \bar{\nu}
    \end{split}
    \label{pa}
\end{equation}
The following scattering processes do not contribute to CFI in the current setup.\\
The neutrino-lepton scattering interaction:
\begin{equation}
    \begin{split}
        \quad\nu + e^- \longleftrightarrow  \nu + e^-\\
        \quad\nu + e^+ \longleftrightarrow  \nu + e^+\\
        \quad\nu + \mu^- \longleftrightarrow  \nu + \mu^-\\
        \quad\nu + \mu^+ \longleftrightarrow  \nu + \mu^+
    \end{split}
    \label{nls}
\end{equation}
The neutrino-nucleon scattering interactions:
\begin{equation}
    \begin{split}
        \text{neutron scattering}\quad\nu+n\longleftrightarrow \nu+n\\
        \text{proton scattering}\quad\nu+p\longleftrightarrow \nu+p
    \end{split}
    \label{nns}
\end{equation}
Among all those interactions, only the processes~\ref{pa} contribute to the collision rates of $\barparent{\nu}_\tau$ in the evaluation of CFI. On the other hand, the scattering processes, Eqs.~\ref{nls} and \ref{nns}, do not contribute to CFI under the assumption of isotropic distribution in momentum space but only contribute to neutrino trapping. In processes involving nucleons (\ref{ccr}, \ref{nd}, and \ref{nns}), the effects of weak magnetism, recoil, and momentum transfer dependence are taken into account. To evaluate the growth rate of CFI, we first compute the energy-dependent inverse mean-free path $1/\lambda_{\nu_i,j}(E)$ and the emissivity $j_{\nu_i,j}(E)$, where $i$ labels the (anti-)neutrino flavor and j labels the relevant interactions. The sum of the rates due to the relevant interactions are the energy-dependent collision rates for the evaluation of CFI parameters (see Eq.~\ref{nGM} and the description there):
\begin{equation}
\Gamma_{\nu_i}(E)=\sum_j\left(1/\lambda_{\nu_i,j}(E)+j_{\nu_i,j}(E)\right).
\end{equation}
The numerical scheme to evaluate these interaction rates is the same as in \cite{10.1093/ptep/ptac118}.

\end{document}